\documentclass[aps,prl,twocolumn,groupedaddress]{revtex4-2}

\usepackage{graphicx}
\usepackage{xcolor}
\usepackage{tcolorbox}

\begin{document}

\title{ 
Low-symmetry polymorph of GaP upends bonding paradigms of metallic high-pressure III-V compounds
}

\author{Barbara Lavina}
\email[]{lavina@uchicago.edu}
\affiliation{University of Chicago, GeoSoilEnviro Center for Advanced Radiation Sources, Chicago, IL, 60439, USA} 
\affiliation{X-ray Science Division, Argonne National Laboratory, Lemont, IL, 60439, USA}

\author{Enrique Zanardi}
\author{Andr\'es Mujica}
\email[]{amujica@ull.edu.es}
\affiliation{Departamento de F\'isica and Instituto Universitario de Materiales y Nanotecnolog\'ia, 
             MALTA Consolider Team, 
	     Universidad de La Laguna, San Crist\'obal de La Laguna 38200, Tenerife, Spain}

\author{Hyunchae Cynn}
\affiliation{High-Pressure Physics Group, Physics and Life Sciences, Lawrence Livermore National Laboratory, Livermore, CA, 94550, USA}

\author{Yue Meng}
\email[]{Current contact: mengyue.cc@gmail.com}
\affiliation{X-ray Science Division, Argonne National Laboratory, Lemont, IL, 60439, USA}

\author{Vitali Prakapenka}
\affiliation{University of Chicago, GeoSoilEnviro Center for Advanced Radiation Sources, Chicago, IL, 60439, USA}

\author{Jesse S. Smith}
\affiliation{HPCAT, X-ray Science Division, Argonne National Laboratory, Lemont, IL, 60439, USA }

\date{\today}

\begin{abstract}

The pressure-induced polymorphism of binary octect compounds has long been
considered a settled problem although the possible atomic disordering of some
phases remains a puzzling observation.
Taking GaP as a case study, we conclude, through x-ray microdiffraction and
first-principles calculations, that its high-pressure phase II
(previously reported as being disordered) adopts in fact an ordered
base-centered monoclinic structure previously unknown in this class of
compounds.
The formation of layered 
patterns with variable degrees of interlayer
dimerization, as observed in GaP, marks a paradigm shift of our understanding
of ordering in octect high-pressure phases which calls for a more extensive
re-examination. 
A rich polymorphism with fine tuning of chemical and physical properties can be
envisioned.

\end{abstract}

\maketitle

Under ambient conditions GaP crystallizes in the zincblende structure (zb)
common to several other III-V and II-VI binary compounds and is an indirect-gap
semiconductor.  A pressure-induced structural transition around 22 GPa, which
is associated with a change towards a semimetallic state, was first reported in
the 1970s \cite{Homan:1975} but the structure of the high-pressure phase GaP-II
remained elusive for more than twenty years.  Taking advantage of interim
experimental advances in high-pressure crystallography, Nelmes, McMahon, and
coworkers re-evaluated in the 1990s the structure of the high-pressure phases
of the III-V and II-VI families unveiling several underlying systematic
behaviors which have prevailed up to now \cite{Nelmes:1998}.  
For GaP, their angle-dispersive powder diffraction experiments showed that the phase obtained
upon ambient-temperature compression could be described as a distorted NaCl
arrangement \cite{Nelmes:1997}, named \emph{Cmcm} after its space group, ruling
out the occurrence of the previously proposed $\beta$-Sn-type structure
\cite{Baublitz:1982}.  Diffraction data are consistent with  
intra-crystalline atomic disorder \cite{Nelmes:1997,Aquilanti:2007} whereas
x-ray absorption spectroscopy data suggest that the phase is ordered in the short
range \cite{Aquilanti:2007}.  As per the established systematics, the
\emph{Cmcm} phase appears ubiquitously in the phase diagram of the III-V and
II-VI families \cite{Nelmes:1998,Mujica:2003}. 

Recently, a new polymorph of GaP was  synthesized at $\sim$43 GPa and high temperature \cite{Lavina:2018}.  It adopts the so-called super-\emph{Cmcm} structure (or \emph{oS}24 in Pearson notation)  similar to that previously
described in InSb \cite{Nelmes:1995a}, but in GaP the phase forms single homonuclear P-P bonds  in striking contrast
with all previous understanding of bonding in the high-pressure polymorphism of
III-V compounds. Finally, the sc16 modification of GaP was found to be stable in a narrow range between 17-20 GPa \cite{Lavina:2022}.
Considering that \emph{oS}24 was synthesized at pressures significantly greater than the onset of metallization, and exhibited instability upon decompression within its calculated stability field  \cite{Lavina:2018}, we suspected that yet another phase between sc16 and \emph{oS}24 might occur. If the dimerization observed in \emph{oS}24   occurs only in a higher-pressure modification of the first metallic phase, it would not be necessarily  related to the physical property change. Therefore we explored such pressure range  to establish  the nature of the first metallic phase of GaP. In addition, we performed extensive theoretical analysis to understand more deeply the evolution of bonding across  polymorphs.

High-pressure and temperature conditions were generated using a laser-heated
diamond anvil cell (DAC). Structures were probed \emph{in situ} with
synchrotron microdiffraction techniques. In order to sample a sufficiently
large portion of  reciprocal space, we used a wide-opening DAC equipped with
conical anvils, providing 64$^\circ$ 4$\theta$ access. Between the anvils'
culets, 400 $\mu$m in diameter, Re gaskets indented to a thickness of about 40
$\mu$m and with laser-drilled holes of 230 $\mu$m diameter  
provided the sample chambers.  A chip or powdered sample (99.999\% purity  from Alfa Aesar) along with ruby and platinum as pressure gauges
\citep{MAO:1976ay,Dewaele:2004uq} were placed in the sample chamber before
filling it with pre-pressurized neon \citep{Rivers:2008}. High temperatures
were obtained using  online double-sided IR laser-heating systems. The
temperature was determined from the emitted radiation \citep{Meng:2015}. The
structure was probed using a microfocused hard x-ray beam at Sector 16
(16-ID-B), HPCAT, and at Sector 13 (13-ID-D), GSECARS, of the Advanced Photon
Source, Argonne National Laboratory. Diffracted beams were collected with either 
MAR165-CCD (16-ID-D) or  Pilatus 1M (13-ID-D) detector that were calibrated using the
powder diffraction patterns of the CeO$_{2}$ and LaB$_{6}$ standards
\citep{Prescher:2015aa}. SXD data, with a resolution down to 0.6 \AA, were
collected with the rotation method and processed with CrysAlis Pro 
\cite{crys} and Shelxl \citep{Sheldrick:2008uq}.

In order to assess the energetics and stability of the new polymorph relative
to the established phases in the phase diagram of GaP, we performed total
energy calculations for this material including the novel phase described in
this Letter and phases previously observed or considered for the III-V family
\cite{Mujica:2003}.  The results that we will show here were obtained within
the \emph{ab initio} framework of the density functional theory (DFT), using a
basis set of plane waves, with the outermost $s$ and $p$ valence electrons of
Ga and P, as well as the Ga semicore $d$ electrons, explicitly dealt with in a
projector augmented wave scheme \cite{Blochl:1994kx,*Kresse:1999aa} as
implemented in the \textsc{vasp} code \cite{Kresse:1993aa,*Kresse:1996fk}; the
PBE generalized gradient approximation (GGA) was adopted for the
exchange-correlation functional \cite{Perdew:1996,*Perdew:1996_erratum}, and
the cutoff in the kinetic energy of the plane waves basis was set at 370 eV
\cite{Note:1,Perdew:2008aa,*Perdew:2008aa_erratum,Giannozzi_QE:2009}.  Dense
{\bf k}-point grids were used for the Brillouin zone integrations, e.g. a
16$\times$16$\times$8 grid was used for the novel \emph{mS}16 phase, which is
metallic with eight atoms within the primitive unit cell (see later).  All the
structures were fully relaxed under hydrostatic conditions through the
calculation of the stress tensor and the forces on the atoms, to residual
stress anisotropies and forces less than 0.1 GPa and 0.005 eV/\AA,
respectively, which ensures a convergence in the relative energies of the
phases at the chosen level of approximation of the order of the meV per formula
unit (pfu).  The phonons for the main phases under discussion were studied using the
small-displacements method, and the effect of temperature was taken into
account at the level of the quasi-harmonic approximation
\cite{Togo:2015,Alfe:2009}.  A fuller account will be given elsewhere \cite{Zanardi:unpub}.

\begin{figure} 
\includegraphics[width=1\columnwidth]{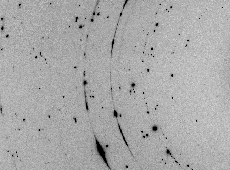}
\caption{\label{fig:data} 
        Example of diffraction data of coarse-grained \emph{mS16}-GaP.
	}
\end{figure}

\begin{figure} 

\includegraphics[width=1\columnwidth]{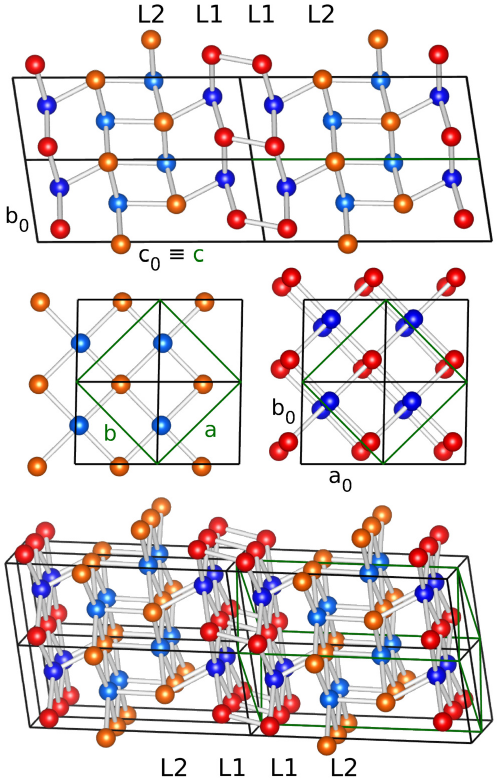}
\caption{\label{fig:structure_mS16} 
        (Top) 
	The \emph{mS}16 structure viewed from the ${a_0}$ axis of the
	primitive unit cell used in the calculations.  Ga1 and P1 sites (within
	the L1 layers, viewed here in profile) are respectively shown in dark
	blue and red, whereas a lighter shade is used for the Ga2 and P2 sites
	that make the L2 layers.
	(Middle)
	View along the $c$ axis showing two adjacent, dimerized L1 layers
	(left) and two adjacent, overlapped L2 layers (right). 
	(Bottom)
	Same as (Top) but showing a 2$\times$2$\times$2 slab with a modicum of
perspective so as to reveal the full connectivity among sites.  In all
these plots the profiles of the primitive-cell boundaries used in the
calculations $({a_0}, {b_0}, {c_0})$ are shown in black
whereas those of the conventional cell $({a}, {b}, {c})$
are shown in green.  (See also further plots in the Supplemental
Material.)
	}
\end{figure}

\begin{table}
\caption{\label{tab:table1}
	Experimental structural data for the \emph{mS}16 phase, corresponding
	to $p$ = 30.7 GPa and $T$ = 300 K.  The structural refinement was
	conducted against 180 measured structure factors, 103 of which unique.
	A total of 13 parameters were refined, including eight fractional
	coordinates, four isotropic displacement parameters ($U_{eq}$) and the
	scale factor. Structure factors statistics is satisfactory with
	$R_{int}$=0.044 and $R\sigma$=0.028. The single-crystal structural
	refinement converged with $R_{4\sigma}$=0.038, $R_{all}$=0.045,
	$wR_2$=0.11.  For comparison, our calculated theoretical values of the
	structural parameters are given in square brackets. 
} 
\begin{ruledtabular}
\begin{tabular}{cccccccc}
\emph{mS}16 & & \multicolumn{5}{l}{SG \emph{C2/m}, No. 12} 
\\
\hline
			    &  $a$ (\AA)  &  $b$ (\AA)  &  $c$ (\AA)  &    &  $\beta$ ($^\circ$)  \\
			    &  4.931(3)   &  4.7189(11) &  9.553(2)   &    &  102.67(3)  \\
			    & [4.905]     & [4.732]     & [9.590]     &    & [102.97]   \\
\hline
 	Sites &    & \multicolumn{4}{c}{Wyckoff positions}    & $U_{eq}$ (\AA$^{2}$) 
\\
\hline
	Ga1 & 4\emph{i} & \multicolumn{4}{c}{0.0443(4), 0, 0.36638(11)}  & 0.0119(9) \\
	    &           & \multicolumn{4}{c}{[0.0408, 0, 0.3662]}        \\
	Ga2 & 4\emph{i} & \multicolumn{4}{c}{0.2504(4), 0, 0.11899(10)}  & 0.0115(9)  \\
	    &           & \multicolumn{4}{c}{[0.2504, 0, 0.1190]}        \\
	P1  & 4\emph{i} & \multicolumn{4}{c}{0.5549(8), 0, 0.3935(3)}    & 0.0142(8) \\
	    &           & \multicolumn{4}{c}{[0.5504, 0, 0.3934]}        \\
	P2  & 4\emph{i} & \multicolumn{4}{c}{0.7510(9), 0, 0.1347(3)}    & 0.0146(8) \\
	    &           & \multicolumn{4}{c}{[0.7516, 0, 0.1342]}        \\
\end{tabular}
\end{ruledtabular}
\end{table}

We performed several  synthesis experiments between  15 and 30 GPa. A new phase was observed between about 20 and 30 GPa. Diffraction images showed that the samples developed variable grain size after
laser-heating (Fig.  \ref{fig:data}).  Single-crystal structural analysis was performed on the largest grains
harvested from diffraction maps of the sample chambers \citep{Lavina_Jove}. At about 31 GPa a monoclinic lattice with $a$=4.931(3) \AA, $b$=4.7189(11) \AA, $c$=9.553(2) \AA,
and $\beta$=102.67(2)$^\circ$  was readily identified. The
lattice showed obvious geometric relations with that of the recently
characterized \emph{oS}24 polymorph \citep{Lavina:2018}, which measured
$a$=4.621 \AA, $b$=4.927 \AA, $c$=13.335 \AA \ at 43 GPa.  Taking into account
the pressure difference, the lengths of the $a$ and $b$ axes of the two
lattices are close, and the projection of the $c$ parameter of the new phase
onto the perpendicular to its $ab$ plane is roughly 2/3 that of the \emph{oS}24
phase. We inferred that the two phases likely have a similar atomic arrangement
in the $ab$ plane, \emph{viz.} a distorted NaCl-like pattern, and that the
monoclinic unit cell of the new phase contains four NaCl-like layers (compared
to six in \emph{oS}24), resulting in a crystallographic unit-cell content of 16
atoms. With such constraints, solving the crystal structure was rather
straightforward, nonetheless the correct space group, \emph{C2/m}, was assigned
after calculations. Single-crystal structural refinements were performed
against the structure factors of selected grains showing the best statistical
parameters for observed intensities, and a selected result is shown in Table
\ref{tab:table1}.  The new polymorph will be hereafter referred to using the
Pearson notation as \emph{mS}16. To the best of our knowledge, this structural
type has not been previously identified among the high-pressure polymorphs of
semiconductors and is a new structure type.  Plots of the \emph{mS}16 structure
are shown in Figure \ref{fig:structure_mS16}.

Two distinct features appear prominently in both the \emph{mS}16 and
\emph{oS}24 phases: (i) their underlying two-dimensional NaCl-like
arrangements, and (ii) the partial dimerization of the phosphorus atoms.  The
structures of both \emph{mS}16 and \emph{oS}24 can be described as different
stackings of layers corresponding to distorted NaCl-(001) planes, of which two
distinct types can be found in both phases: L1 layers, hosting the dimerizing
P1 phosphorus and the Ga1 atoms; and L2 layers, containing the non-dimerizing
P2 phosphorus and the Ga2 atoms (see Fig.  \ref{fig:structure_mS16}).  The
P1-P1 dimers lie across adjacent L1 layers in both polymorphs.  The stacking is
different than in NaCl as like atoms nearly overlap (anti-NaCl).  The L1 layer
buckling is the result of two nearest P1 atoms approaching to form the dimer,
whereas the two nearby Ga1 atoms maintain a greater distance.  In \emph{mS}16
there are also two adjacent L2 layers, and their relative stacking, while
distorted, is NaCl-like, with interlayer L2-L2 bonds directed along the $c$
direction, hence all closest neighbors are unlike atoms both within and across
the pair of L2 layers.  The stacking of L1 and L2 adjacent layers is shifted
with respect to the NaCl structure by roughly 1/4-unit cell along the $a$-axis
for \emph{mS}16 and along the $b$-axis for \emph{oS}24.  The resulting
stacking sequence is L1-L1-L2-L2 in \emph{mS}16 and L1-L1-L2-L1-L1-L2 in
\emph{oS}24, so that half the P atoms are dimerized in \emph{mS}16 whereas 2/3
form monatomic bonds in \emph{oS}24.  As an effect of the interlayer
dimerization, the L1 layers exhibit a slightly larger amount of corrugation
than the L2 layers, as can be seen in Fig. \ref{fig:structure_mS16}.  In
\emph{mS}16 there is no puckering of the atomic rows \emph{within} the
NaCl-like layers (either L1 or L2), only transverse (out of plane)
buckling/corrugation.  This is unlike \emph{oS}24 
(and \emph{Cmcm} \cite{Nelmes:1998}) for which
there is puckering within the L2 layers whereas, conversely, these layers do
not have any corrugation -- they are perfectly flat, see Supplemental Material.
The intra-layer puckering of \emph{oS}24 is related to the reduced coordination
of the P2 sites in this phase \cite{Note:2}.

\begin{figure} 
\includegraphics[width=1\columnwidth]{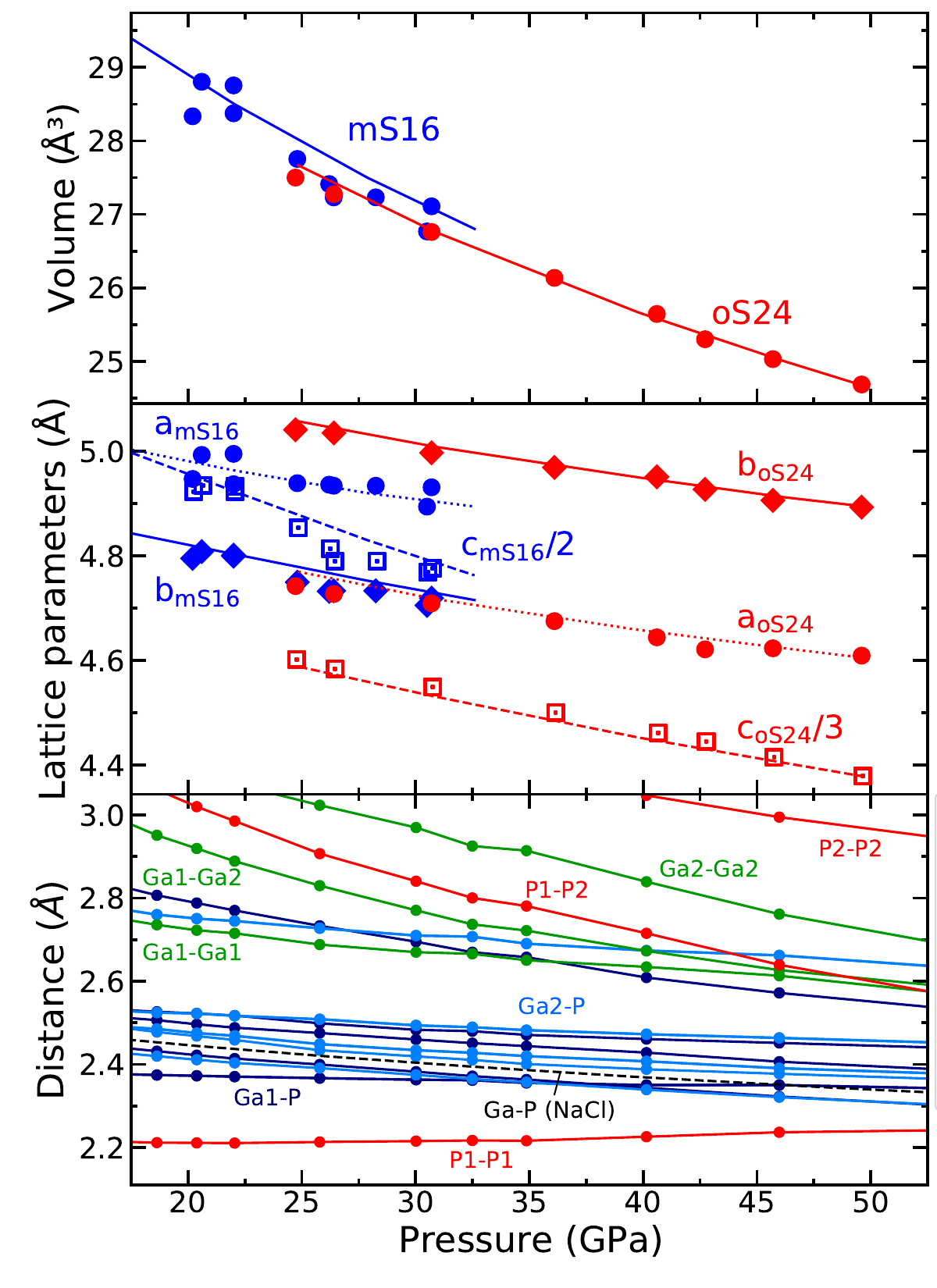}
\caption{\label{fig:lattP}
	Comparison of the pressure evolution of the unit-cell geometry of
	\emph{mS}16 and \emph{oS}24: Volume per formula unit (top panel) and
	$a$, $b$, and $c$ lattice parameters (central panel).  
	Blue symbols represent experimental \emph{mS}16 data, whereas red
	symbols show \emph{oS}24 data, and the lines correspond to calculated
	values.  The $c$ parameters of \emph{mS}16 and \emph{oS}24 are shown
	respectively divided by 2 and 3.  
	Bottom panel: Calculated pressure evolution of near-neighbour distances
	in \emph{mS}16.  As a reference, the nearest-neighbor Ga-P distance for
	the ideal NaCl phase is also provided (dashed black curve).}
\end{figure}

The likeness of \emph{mS}16 and \emph{oS}24 is further evidenced in their
similar densities (Fig. \ref{fig:lattP}, top).  The $b$-axis of \emph{mS}16 and
the $a$-axis of \emph{oS}24 lie approximately on the same compression curve
(Fig. \ref{fig:lattP}, center), reflecting the stacking correspondence in these
directions between the two phases. The $a$-axis of \emph{mS}16 is
systematically smaller than the $b$-axis of \emph{oS}24.  At any given
pressure, \emph{mS}16 shows a larger average thickness of the NaCl-like layers,
the greater the degree of dimerization the larger is the $ab$ plane and the
shorter is the average distance between layers.

\begin{figure}
\includegraphics[width=1\columnwidth]{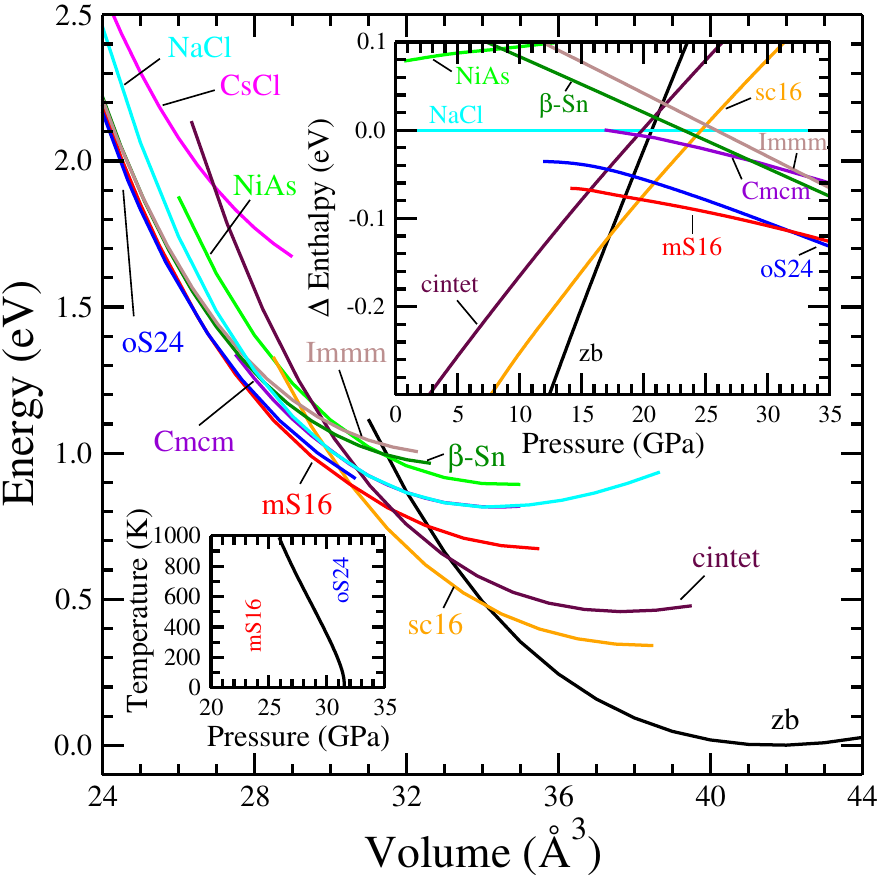}
\caption{\label{fig:EV} Energy-volume relations (per formula unit) for the various calculated structures. The energy is given with respect to the zero-pressure state of the zb phase. In the insets, the enthalpies of the calculated phases (given with respect to that of the NaCl phase) and the $p$-$T$ phase boundary between \emph{mS}16 and \emph{oS}24 are provided.
}
\end{figure}

\begin{figure}
\includegraphics[width=1\columnwidth]{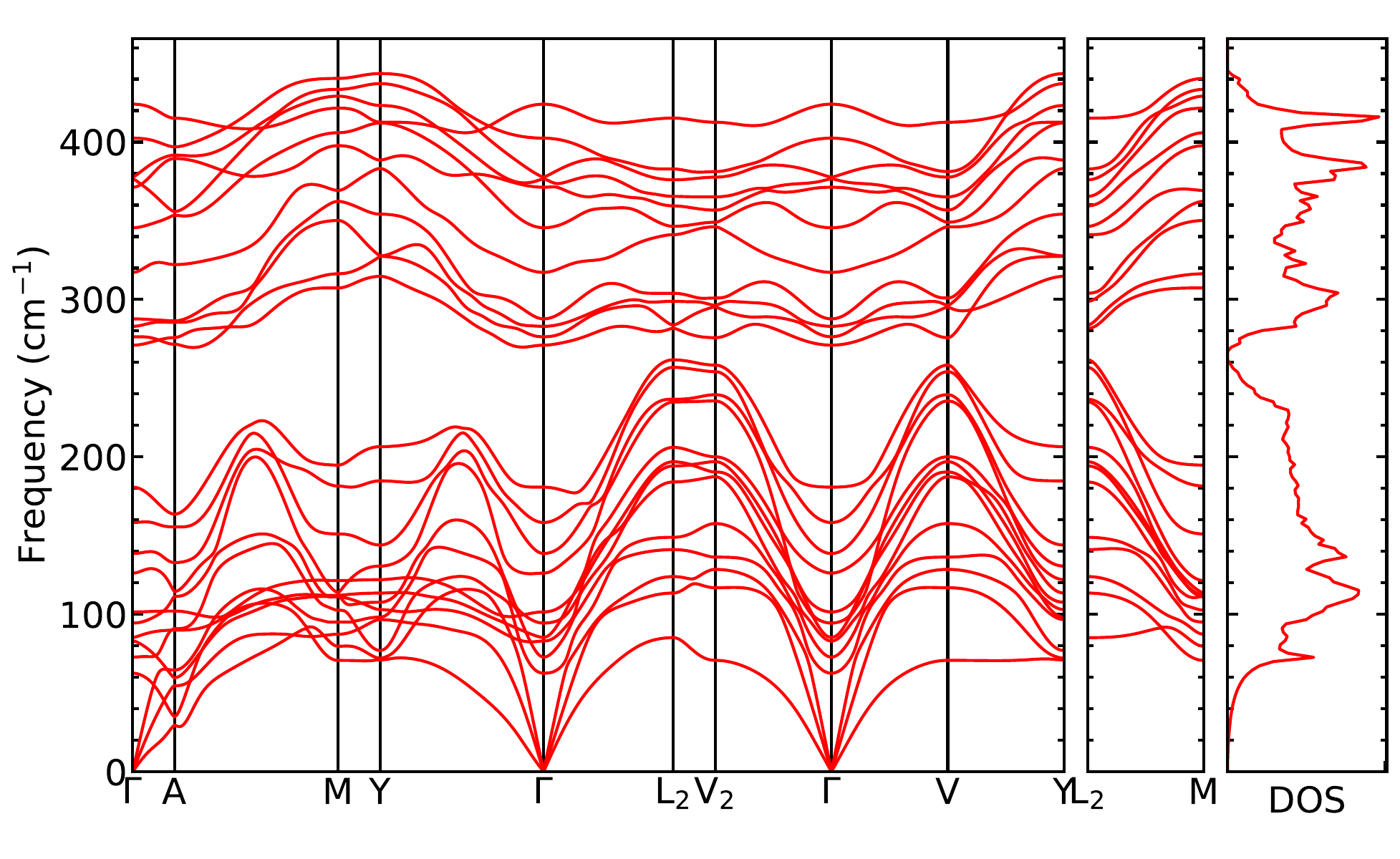}
\includegraphics[width=1\columnwidth]{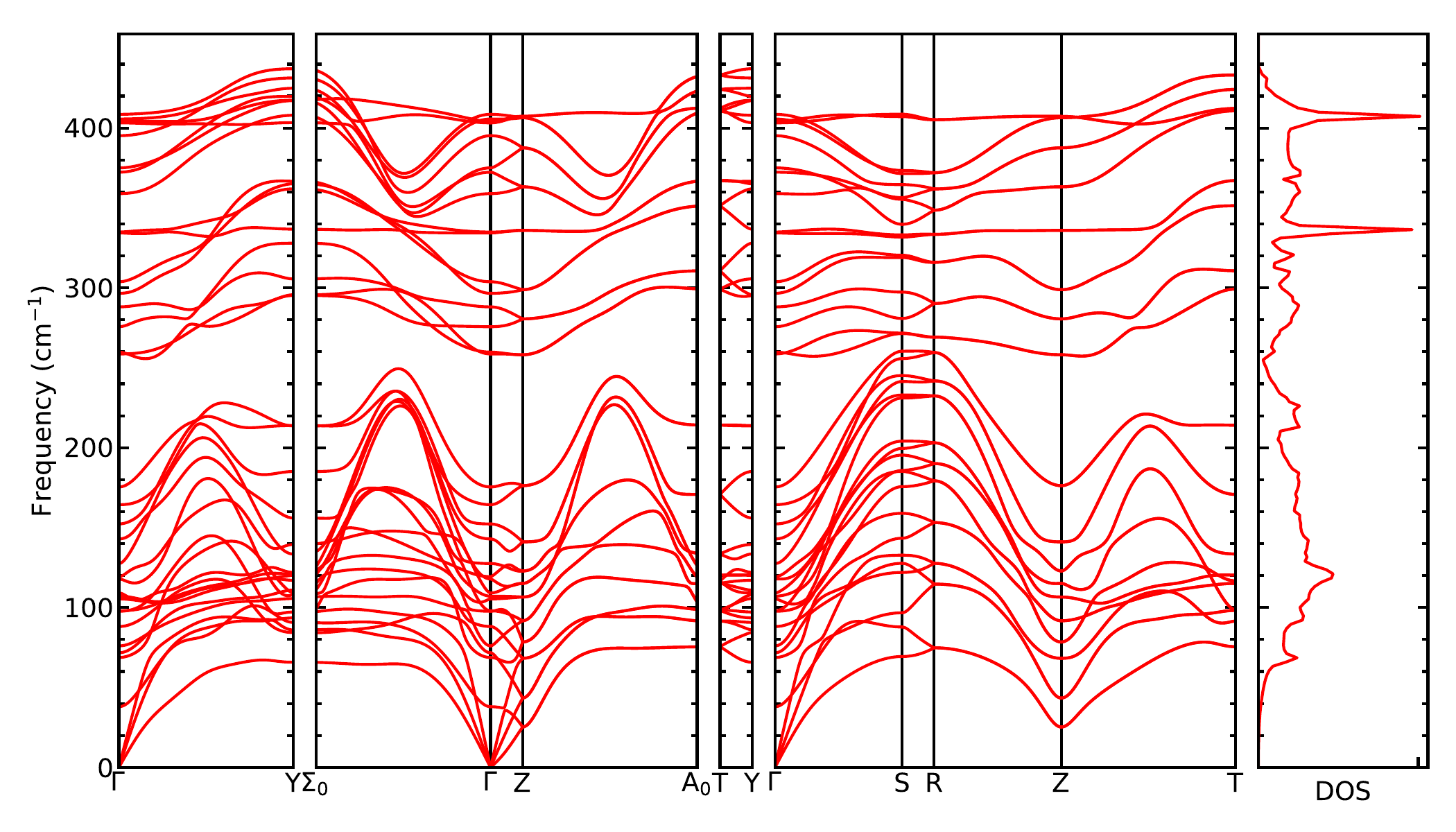}
\caption{\label{fig:PH} 
       (Top) Calculated phonon bands along selected segments within
       the Brillouin zone of the \emph{mS}16 phase at $\sim$32 GPa.  
       (Bottom) Idem for the \emph{oS}24 phase.
	}
\end{figure}

Figure \ref{fig:EV} shows the calculated energy-volume curves, $E(V)$, and
enthalpy-pressure curves, $H(p)$, for the different structural phases of GaP
considered in our \emph{ab initio} DFT calculations, including the novel
\emph{mS}16 phase as well as the \emph{Cmcm}, $\beta$-Sn, \emph{oS}24, and
sc16 phases, among others \cite{Mujica:2003}.  These results show \emph{oS}24
and \emph{mS}16 to be very close in energy/enthalpy (and hence also density) in
the whole range of compressions investigated, with \emph{oS}24 favored at the
highest pressures and, viceversa, \emph{mS}16 favored in the lower pressure
range, which is a qualitative trend consistent with the experimental
observations about the pressure conditions of the synthesis of these two
polymorphs and their intervals of existence.  The calculated
\emph{mS}16/\emph{oS}24 coexistence pressure is around 32 GPa however this
value should be taken \emph{cum grano salis} given the closeness of their
respective $E(V)$ curves which results in a large uncertainty, with temperature
favoring the highest-pressure \emph{oS}24 phase, see inset in Fig. \ref{fig:EV} \cite{Note:3}.  Notwithstanding
this unavoidable uncertainty related to the similarities of the two phases, all
our calculations consistently support the high-pressure stability of
\emph{mS}16 followed by that of \emph{oS}24.  The calculated phonon frequencies
of \emph{mS}16 (see Fig. \ref{fig:PH}) show that it is dynamically stable
at these high pressures.

 As recently reported \cite{Lavina:2022}, a narrow stability range for  the sc16 polymorph is found between zb and \emph{mS}16. 
On the other hand, around  $\sim$30 GPa, the 
enthalpies of the $\beta$-Sn and \emph{Cmcm} phases
lie significantly higher ($\sim$70 meV pfu) than either \emph{mS}16 or
\emph{oS}24, which rules $\beta$-Sn and \emph{Cmcm} out as thermodynamically
stable in the phase diagram of GaP, although they have been previously
considered in experiments \cite{Nelmes:1997,Aquilanti:2007} and theoretical
studies \cite{Mujica:1997,Mujica:1998,Ozolins:1999}.  The \emph{Cmcm} phase can be
understood in terms of a simple sliding pattern of alternate planes from the ideal NaCl
structure \cite{Nelmes:1998}, which relates to an intrinsic TA($X$)
zone-boundary mode softening of the
NaCl-type phases in III-Vs and the ensuing dynamical instability
\cite{Mujica:1996aa,Ozolins:1999}. 
Our calculations are in agreement with this established view, however any possible stability range for \emph{Cmcm}-GaP 
is completely obliterated when the dimerized 
polymorphs are taken into consideration.
As previously described, \emph{mS}16 and \emph{oS}24 can also be viewed in
terms of larger patterns of sliding NaCl planes but while the structural
flexibility of a four-plane (\emph{mS}16) or six-plane (\emph{oS}24)
stacking pattern allows the formation of interlayer dimers, in various
proportions, the simplest two-plane pattern of the \emph{Cmcm} structure does
not allow any such dimerization.  

The \emph{mS}16 phase is characterized by  bonding anisotropy and distorted
coordination geometries (Fig. \ref{fig:structure_mS16}).  The shortest bond of
the structure is that of the P-P dimer, measuring 2.217(6) \AA \ at 30.7 GPa, 
a length typical of a single covalent P-P bond \cite{Pottgen:2011aa}. 
P-Ga interatomic distances (Fig. \ref{fig:lattP}, bottom) are clustered between
$\sim$2.35 and $\sim$2.5 \AA, with a clear gap to longer distances
starting at about 2.7 \AA, thus we  consider bonding interactions all the Ga-P
distances up to about 2.5 \AA .
In addition to the dimer, P1 forms four bonds with Ga1 atoms
measuring between 2.36 and 2.48 \AA, the coordination geometry is strikingly
similar to that of the corresponding atom in \emph{oS}24 \citep{Lavina:2018}.
The non-dimerizing P2 atom is surrounded by six Ga atoms in a distorted octahedral
coordination with bonds lengths ranging between 2.36 and 2.50 \AA. The angles
between P2 - Ga2 bonds within the L2 layer are close to 90$^\circ$, whereas the
angles involving  off-plane  bonds are as small as 67$^\circ$ and
72$^\circ$.  In \emph{oS}24, the corresponding non bonding Ga atom lies at
about 2.6 \AA \ from P2, and while all the angles between P2 and the four
closest Ga atoms are close to 90$^\circ$, the bond angles formed with the
apical Ga measure between 116$^\circ$ and 127$^\circ$, defining a clear
fivefold coordination. Therefore, unlike P1, the coordination geometry of P2
shows some small but significant differences in the two dimerized phases,
resulting in a coordination number six for P2 in \emph{mS}16 and five
for P2 in the higher-pressure phase \emph{oS}24. Hence the two  layered GaP
phases show a rare exception to the general rule of increasing
coordination number with pressure in polymorphs.

The behavior under pressure of the P-P dimer that stems from the \emph{ab
initio} calculations (see Fig.\ref{fig:lattP}, bottom) shows that it is
remarkably rigid within the range of observation of the \emph{mS}16 phase,
whereas its length \emph{increases} slightly at higher pressures, which is
indicative of a slight weakening of the homonuclear bond.  On the other hand, the
various intralayer Ga-P bonds follow, unremarkably, approximately the same
compressibility curve as that calculated for the reference NaCl phase in the
interval of interest.

\begin{figure}
\includegraphics[width=1\columnwidth]{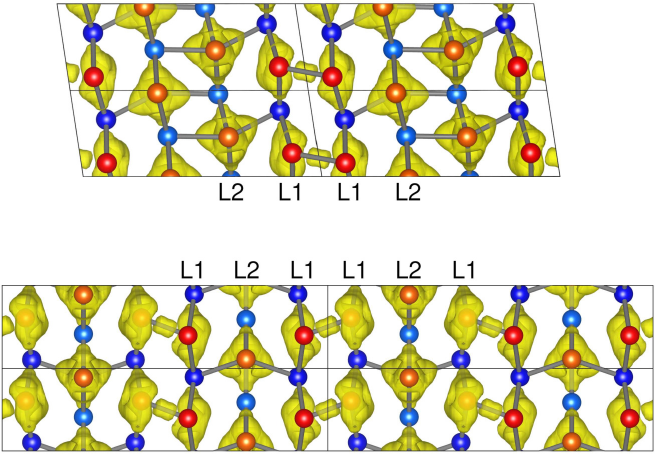}

\caption{\label{fig:ELF} 
	Calculated ELF isosurfaces for the valence $s$ and $p$ electrons of the
	\emph{mS}16 (top) and \emph{oS}24 (bottom) phases, as seen along the
	$a$ axes of their respective primitive cells (outlined in black). 
        The value of the ELF was set at 0.75. 
	}
\end{figure}

Calculated ELF isosurfaces (valence only) for the two dimerized polymorphs are
shown in Fig. \ref{fig:ELF} \cite{Momma_VESTA:2011}.  Once again, these plots
show that there are strong similarities between the two phases
and also subtle differences. 
The covalent P-P bonding is clearly seen in both
phases as well-defined large-ELF domains between nearby P1 atoms. The ELF
around P1 is nearly identical in the two phases, the four Ga--P
bonds produce a doughnut shaped ELF domain in the \emph{ab} plane.  Fig.
\ref{fig:ELF} highlights the subtle differences in ELF around the
non-dimerizing phosphorus in \emph{mS}16 and \emph{oS}24.  In the projection
shown in Fig. \ref{fig:ELF} the ELF shows nearly perfect threefold symmetry
around P2 in \emph{oS}24, such symmetry is broken in \emph{mS}16, and the ELF
domain at 0.75 protrudes in the direction of the sixth Ga atom, supporting the occurrence of a weak bond.

In summary, two decades after the high-pressure polymorphism of the binary
semiconductors was considered a substantially settled matter, our findings
reveal the occurrence of a new structure type, adopted by the first
high-pressure metallic polymorph of gallium phosphide.  We have unraveled a
marked complexity in the metallic phases, arising from strong
bonding differentiation and finely tuned pressure dependence of the degree of
dimerization.  It is worth noting that the work presented here is not a mere
redetermination of the phases observed in previous studies and interpreted as
\emph{Cmcm} \cite{Nelmes:1997,Aquilanti:2007}. Without heating, the complex
ordering of \emph{mS}16 and \emph{oS}24 cannot be achieved (the powder
diffraction pattern would show more peaks than observed in the cited studies).
Small domains locally resembling \emph{mS}16 or
\emph{oS}24 could however form. Hence a distorted NaCl-like phase, locally ordered,
might still be a reasonable description for the metastable disordered material obtained
upon ambient-temperature compression of GaP.

The discovery of \emph{mS}16, along with the previously-described \emph{oS}24
polymorph of GaP, challenges previous views concerning the unfavorable role
played by bonds between like atoms in the energetics of the high-pressure
phases of the III-V and II-VI binary compounds and emphasizes the role of
NaCl-derived layered structures \emph{with} partial interlayer dimerization
as a relevant type of ordering in these compounds.
The role of complex ordering derived from the NaCl basic arrangement
has been minimized \cite{Kelsey:2000aa}, but it is clear that while the
total-energy difference between such polymorphs is small, they are distinct phases
and their occurrence is hereby explained. The layer stacking allows for the
accommodation of variable degrees of phosphorus dimerization. 
Distinct ordering sequences of such layers, and therefore new structures, are
likely to occur in other 
octet compounds as well, an aspect relevant to material design, as
phases and properties might be finely tunable with pressure.
\\
\\
\\

\begin{acknowledgments}

This research was partially supported by COMPRES, the Consortium for Materials
Properties Research in Earth Sciences, under NSF Cooperative Agreement
EAR 1606856.  
AM acknowledges financial support through Projects
MAT2016-75586-C4-3-P (MINECO, Spain), PID2019-106383GB-C43 and PID2022-138076NB-C44 (MCIN/AEI/10.13039/501100011033).  
The MALTA Consolider Team Research Network is supported by Project 
RED2022-134388-T (MCIN, Spain).
Portions of this work work were conducted at GeoSoilEnviroCARS
(GSECARS), Sector 13 and High Pressure Collaborative Access Team
(HP-CAT), Sector 16, 
Advanced Photon Source (APS), Argonne National Laboratory.  
GSECARS is supported by the National Science Foundation - Earth Sciences
(EAR-1634415) and DOE-GeoSciences (DE-FG02-94ER14466). 
HPCAT operations are supported by DOE-NNSA’s Office of Experimental Sciences.  The Advanced Photon Source is a U.S. Department of Energy (DOE) Office of Science User Facility operated for the DOE Office of Science by Argonne National Laboratory under Contract No. DE-AC02-06CH11357 
Lawrence Livermore National Laboratory is operated by Lawrence
Livermore National Security, LLC, for the U.S.  Department of Energy
(DOE), National Nuclear Security Administration under Contract
DE-AC52-07NA27344.  
Use of the COMPRES-GSECARS gas loading system was supported by COMPRES under
NSF Cooperative Agreement EAR-1606856 and by GSECARS through NSF grant
EAR-1634415 and DOE grant DE-FG02-94ER14466.

\end{acknowledgments}

\bibliography{Lavina et al_2024_mS16-GaP
}

\end{document}